\documentclass[11pt]{article}

\title{Explicit RIP Matrices in Compressed Sensing from Algebraic Geometry}

\author{Hao Chen
  \thanks{H. Chen is with the Department of Mathematics, School of Sciences, Hangzhou Dianzi University, Hangzhou  310018, Zhejiang Province, China. H.Chen was supported by NSFC Grant 11371138.}}

\begin{document}

\maketitle

\begin{abstract}
Compressed sensing was proposed by E. J. Cand\'es, J. Romberg, T. Tao, and D. Donoho for efficient sampling of sparse signals in 2006 and has vast applications in signal processing. The expicit restricted isometry property (RIP) measurement  matrices are needed  in practice. Since 2007 R. DeVore,  J. Bourgain et al and R. Calderbank et al have given several  deterministic cosntrcutions of RIP matrices from various mathematical objects.  On the other hand the strong coherence property of a measurement matrix was introduced by Bajwa and Calderbank et al  for the recovery of signals under the noisy measuremnt.  In this paper we propose  new  explicit  construction of real valued RIP measurement matrices in  compressed sensing  from algebraic geometry. Our construction indicates that using more general algebraic-geometric objects rather than curves (AG codes), RIP measurement matrices in compressed sensing can be constructed  with much smaller coherence and  much bigger sparsity orders. The RIP matrices from algebraic geometry also have a nice asymptotic bound matching the bound from the previous constructions of Bourgain et al and the small-bias sets. On the negative side, we prove that  the RIP matrices from DeVore's construction, its direct algebraic geometric generalization and one of our new construction do not satisfy the strong coherence property. However we give a modified version of AG-RIP  matrices  which satisfies the strong coherence property. Therefore the new RIP matrices in compressed sensing from our modified algebraic geometric construction  can be used for the recovery of signals from the noisy measurement.\\


\end{abstract}

\section{Introduction}

Compressed sensing  is a technique that measures a sparse signal in ${\bf x} \in {\bf R}^N$ which has $k$ non-zero coordinates at sampling rates that are substantially much lower than the Nyquits-Shannon rate (\cite{CRT,Candes,CT,TC,Donoho,Candes1,Candes2}). For a measurement matrix ${\bf \Phi}$ which has $n$ budged rows and sampling $k$-sparse signals in ${\bf R}^N$ ($N$ columns, $n<N$),  the measurement  is ${\bf y}={\bf \Phi} \cdot {\bf x}$, where $x \in {\bf R}^N$ is a vector with only at most $k$ non-zero coordinates. We refer to \cite{CRT,Candes1,Donoho} for the effective recovery algorithm of the original signal ${\bf x}$ from ${\bf y}$. We say  that a matrix ${\bf \Phi}$ satisfies the restricted isometry property (RIP) of order $k$ with the constant $\delta_k$ satisfying $0 \leq \delta_k <1$, if for every $k$-sparse vector (that is, only $k$ coordinates of the signal are non-zero) ${\bf x} \in {\bf R}^N$, $1-\delta_k \leq \frac{||{\bf \Phi} \cdot {\bf x}||}{||{\bf x}||}\leq 1+\delta_k$. The RIP of the measurement matrix ${\bf \Phi}$ guarantees the effective recovery of the $k$-sparse signal ${\bf x} \in {\bf R}^N$ from ${\bf y}={\bf \Phi} \cdot {\bf x}$ (\cite{Candes, CRT} via linear programming. Here we should note that the RIP is only the sufficient condition of the recovery (\cite{Candes1,Candes2,CRT}). It has been shown that random matrices satisfy the RIP with high probability. However in practice the sampling has to be done deterministically. Explicit RIP measurement matrices are needed in practical compressed sensing. There have been a lot of deterministic  constructions of RIP measurement matrices from various mathematical objects such as polynomials over finite fields, codes, graphs, and additive combinatorics and number theory( see \cite{Alon, Ben, BDFK, DeVore,CHJ,Amini,AHSC,JXC,HCS,Mohads,LG,NT,YZ}). For further extension of compressed sensing in various practices we refer to \cite{BCM, BCJ, CTY, CLMW, CENR,Candes2}. For the theoretical analysis of the sparse-order $k$ and the measurement budget $n$ we refer to paper \cite{CDD}.\\

For a $n \times N$ matrix ${\bf \Phi}$ with $N$ columns ${\phi}_1, ....,{\bf \phi}_N$, the coherence is  $\mu_{{\bf \Phi}}=max_{i \neq j} \frac{|<\phi_i,\phi_j>|}{||\phi_i|| \cdot ||\phi_j||}$. We have $\mu_{{\bf \Phi}} \geq \sqrt{\frac{N}{n(N-n)}}$ from the Welch bound. It is proved in \cite{BDFK,Candes} that a matrix ${\bf \Phi}$ with the coherence $\mu_{{\bf \Phi}}$ satisfies the RIP with the sparsity order $k \leq \frac{1}{\mu_{{\bf \Phi}}}+1$. Thus it is desirable to give explicit construction of matrices with small coherence in compressed sensing.\\

The average coherence of a $n \times N$  matrix ${\bf \Phi}$ with $N$ columns $\phi_1,...,\phi_N \in {\bf R}^n$  is defined as $\omega_{{\bf \Phi}}=\frac{1}{N-1} max_{1 \leq i \leq N}  \Sigma_{j \neq i} \frac{|<\phi_i, \phi_j>|}{||\phi_i|| \cdot ||\phi_j||}$ (\cite{BCM}). As analysised in \cite{BCM,BCJ}, this parameter is important for the recovery of sparse signals from the measurement in the presence of  noise.   The $n \times N$ measurement matrix ${\bf \Phi}$ satisfies the strong coherence property if  $ \mu_{{\bf \Phi}} \leq \frac{1}{160log N}$ and $\omega_{{\bf \Phi}} \leq \frac{\mu_{{\bf \Phi}}}{\sqrt{n}}$. The main result of section 2.2 of \cite{BCM} gave an algorithm for the recovery of sparse signals from noisy measurement provided that the measurement matrix has the strong coherence property.\\

The first systematic deterministic construction of RIP matrices is due to R. DeVore \cite{DeVore}. R. Calderbank and B. Hassibi and their collaborators gave many deterministic constructions of RIP (or statistical isometry property)  matricse in \cite{AHSC,JXC,HCS,CTY}. In 2011 J. Bourgain, S. Dilworth, K.Ford, S. Konyagin and D. Kutzarova \cite{BDFK} gave explicit RIP matrices satisfying $n=o(k^2)$ where $n$ is the row size and $k$ is the sparse order,  from new estimates about exponential sums and additive combinatorics. Their $n \times N$ RIP matrices have coherence $\mu$ matching up the asymptotic bound $\mu=O((\frac{logN}{nlog(n/logN)})^{1/3})$ in the range $log N \leq n \leq (logN)^4$ (page 149 of \cite{BDFK}). From the construction in \cite{Ben} RIP $n \times N$ matrices with size parameters in the range $(logN)^{2.5} \leq n \leq (logN)^5$ and the coherence $\mu=O((\frac{logN}{n^{4/5}loglogN})^{1/2})$ can be constructed. There have been many explicit constructions of RIP matrices from various mathematical objects such as chirp sensing codes (\cite{AHSC}), BCH codes (\cite{Amini}), Reed-Muller codes ( \cite{HCS}), orthogonal codes (\cite{YZ}), Reed-Solomon codes (\cite{Mohads}) and expander graphs (\cite{JXC}). Some early works for other motivations  \cite{Alon,Ben,BDFK} led to RIP matrices with small coherence. The asymptotic behaviour of the coherences of RIP matrices from the constructions in \cite{Ben,BDFK} are nice (see \cite{BDFK}, pages 148-149). However the further strong coherence property of these RIP matrices has not been considered in these previous constructions.\\

The basic construction of \cite{DeVore} is as follows. For a polynomial $f$ with degree less than or equal to  $r-1$ in ${\bf F}_p[x]$ where ${\bf F}_p$ is a finite field with $p$ elements (here $p$ is a prime number, it can also be used for finite field with $q=p^t$ elements), the length $p^2$ vector $v_f=(f_{(a,b)})$ is determined by its $p^2$ coordinates $f_{(a,b)}$ for $(a, b) \in {\bf F}_p \times {\bf F}_p$. Here $f_{(a,b)}=0$ if $f(a) \neq b$, $f_{(a,b)}=1$ if $f(a)=b$. Then the $p^r$ columns of these  lenght $p^2$ vectors give a $p^2 \times p^r$ matrix. It was proved in [4] that the coherence of this $p^2 \times p^r$ matrix ${\bf D}$ satisfies that $\mu_{{\bf D}} \leq \frac{r-1}{p}$.\\

This construction has been generalized to an arbitrary projective non-singular algebraic curve (see \cite{LG}). Let  ${\bf X}$ be a projective non-singular algebraic curve defined over a finite field ${\bf F}_q$ of genus $g$,   ${\bf P}=\{P_1,...,P_{|{\bf P}|}\}$ be a set of ${\bf F}_q$ rational points on the curve ${\bf X}$ and ${\bf G}$ be a ${\bf F}_q$ rational divisor. The functions are taken from the function space $ L({\bf G})$  associated with a ${\bf F}_q$ rational divisor ${\bf G}$. For each $f \in L({\bf G})$, the length $q |{\bf P}|$ vector ${\bf v}_f=(f_{(a,b)})$ where $(a,b) \in {\bf F}_q \times {\bf P}$,  is defined as follows. $f_{(a,b)}$ is $0$ if $f(b) \neq a$ and $f_{a,b)}$ is $1$ if $f(b)=a$. Thus the number of the columns is $dim(L({\bf G}))$ which can be computed from the Riemann-Roch theorem ([12, 17]). We have $dim(L({\bf G}))=deg{\bf G}-g+1$ if $deg{\bf G} \geq 2g$ is satisfied. Each function $f \in L({\bf G})$ leads to a lenghth $q |{\bf P}|$ binary vector $v_f$. This $q|{\bf P}| \times dim(L({\bf G})) $ matrix has its coherence $\mu \leq \frac{deg ({\bf G}) }{|{\bf P}|}$ (see \cite{LG,TV}).\\

In this paper we give several constructions of  explicit RIP matrices  with small coherence from general projective algebraic varieties over finite fields. The calculation of coherence is based on the counting of rational points of these projective algebraic varieties. Thus some results about the estimation of the number of rational points of projective algebraic varieties play an important role. Then some examples of explicit RIP matrices from Fermat surface, projective spaces, ruled surface, Deligne-Lusztig surface and tori surfaces are given. The RIP matrices from our construction C can match  the asymptotic bounds of the RIP matrices constructed in J. Bourgain et al {\cite{BDFK}) and \cite{Ben} in some range. Generally speaking from our new algebraic geometric construction RIP measurement matrices with much smaller coherence and then much bigger sparsity orders than that of the previously constructions can be obtained. On the negative side, we prove that the RIP matrices from the DeVore construction and its direct generalization in \cite{LG} have no strong coherence property. However we present a modified version of our constructsion C (including previous DeVore and AG-code construction in \cite{LG}) which leads to RIP matrices satsfying the strong coherence property. This shows that explicit RIP matrices from algebraic geometry have small coherence and can be used in compressing sensing even in the sampling in the presence of noisy.\\

\section{RIP matrices with small coherence from general projective algebraic varieties}

\subsection{Construction A}
Let ${\bf X}$ be a projective non-singular algebraic curve defined over the finite field ${\bf F}_q$ with $q$ elements, where $q$ is a prime power. Let ${\bf G}=G_1+\cdots+G_t$ be a ${\bf F}_q$-rational  divisor on ${\bf X}$ and ${\bf P}=P_1+\cdots+P_{|{\bf P}|}$ be a set of $|{\bf P}|$ ${\bf F}_q$ rational points on ${\bf X}$. For every rational function in the function space $L({\bf G})$, that is, every rational function on ${\bf X}$ with at most order 1 poles at the points $G_1, .., G_t$. We associated a length $(q+1) (t+|{\bf P}|)$ vector ${\bf v}_f=f_{(a,b)}$ to $f$, where $a \in {\bf F}_q \cup \{@\}, b \in \{G_1,...,G_t\} \cup \{P_1,...,P_{|{\bf P}|}\}$. $f_{(a,b)}=0$ if $f(b) \neq a$, $f_{(a,b)}=1$ if $f(b)=a$ and $a \neq @$ and $f_{(a,b)}=-1$ if $f(b)=a=@$ and $b$ is a pole point of the function $f$. Then we have a $(q+1)(t+|{\bf P}|)\times q^l$ matrix ${\bf \Phi}_{{\bf G}, {\bf P}}$ where $l=dim(L({\bf G}))$ is the dimension of the function space which can be computed from the Riemann-Roch theorem.\\

{\bf Theorem 2.1.} {\em The coherence of this matrix $\mu_{{\bf \Phi}_{{\bf G},{\bf P}}} \leq \frac{2deg({\bf G})}{deg({\bf G})+|{\bf P}|}$.}\\

{\bf Proof.} For any function $f \in L({\bf G})$, the length $(q+1)(t+ |{\bf P}|)$ vector $v_f$ is non-zero only at the points $(f(P_1),P_1),...,(f(P_{|{\bf P}|}), P_{|{\bf P}|})$ and $(@, G_{i_1},...,(@,G_{i_{t_1}})$ where $G_{i_1},...,G_{i_{t_f}}$ are the poles of this function $f$. Thus $|<v_f,v_g>| \leq deg{\bf G}+min\{t_f,t_g\}$. Here $t_f$ and $t_g$ are the numbers of pole points of functions $f$ and $g$. On the other hand $||v_f||=\sqrt{|{\bf P}|+t_f}$ and $||{\bf v_g}||=\sqrt{|{\bf P}|+t_g}$.  From an easy computation in calculus the coherence $\frac{|<v_f,v_g>|}{||v_f||\cdot ||v_g||}$ attains its upper bound only when $f$ and $g$ have 1st pole at all $t$ points $G_1,...,G_t$. Then $t_f=t_g=t=deg{\bf G}$ and $|<v_f,v_g>|=deg{\bf G}+t=2deg{\bf G}$. In this case $||v_f||=||v_g||=\sqrt{|{\bf P}|+deg{\bf G}}$.\\

We can also get another real valued RIP measurement matrices as follows. Let ${\bf X}$ be a projective non-singular algebraic curve defined over the finite field ${\bf F}_q$. Let ${\bf G}=tG$ be a ${\bf F}_q$-rational  divisor on ${\bf X}$ where $G$ is a ${\bf F}_q$ rational point of ${\bf X}$ and $t$ is positive integer. Let  ${\bf P}=P_1+\cdots+P_{|{\bf P}|}$ be a set of $|{\bf P}|$ rational points on ${\bf X}$. For every rational function in the function space $L({\bf G})$, that is, every rational function on ${\bf X}$ with at most order $t$ pole at the point $G$. We associated a length $(q+1) (1+|{\bf P}|)$ vector ${\bf v}_f=f_{(a,b)}$ to $f$, where $a \in {\bf F}_q \cup \{@\}, b \in \{G\} \cup \{P_1,...,P_{|{\bf P}|}\}$. $f_{(a,b)}=0$ if $f(b) \neq a$, $f_{(a,b)}=1$ if $f(b)=a$ and $b \neq G$,  $f_{(a,b)}=-m$ if $b=G$ and $f$ has a order $m$ pole at the point $G$. Then we have a $(q+1)(1+|{\bf P}|) \times q^l$ matrix ${\bf \Phi}_{{\bf G}, {\bf P}}$. Here $l=dim(L({\bf G}))$ is the dimension of the function space which can be computed from the Riemann-Roch theorem.\\

{\bf Theorem 2.2.} {\em The coherence of this matrix $\mu_{{\bf \Phi}_{{\bf G},{\bf P}}} \leq \frac{deg({\bf G})+deg({\bf G})^2}{|{\bf P}|+deg({\bf G})^2}$.}\\

{\bf Proof.}  For any function $f \in L({\bf G})$, the length $(q+1) \times (1+|{\bf P}|)$ vector $v_f$ is non-zero only at the points $(f(P_1),P_1),...,(f(P_{|{\bf P}|}),P_{|{\bf P}|})$ and $(@, G)$. Thus $|<v_f,v_g>| \leq deg{\bf G}+t_ft_g$ where $t_f$ and $t_g$ are the pole orders of the functions $f$ and $g$ at the point $G$ . On the other hand $||v_f||=\sqrt{|{\bf P}|+t_f^2}$ and $||{\bf v_g}||=\sqrt{|{\bf P}|+t_g^2}$.  From an easy computation in calculus the coherence $\frac{|<v_f,v_g>|}{||v_f||\cdot ||v_g||}$ attains its upper bound only when $f$ and $g$ have $t$ order pole at the point $G$. Then $t_f=t_g=t=deg{\bf G}$ and $|<v_f,v_g>|=deg{\bf G}+t^2$. In this case $||v_f||=||v_g||=\sqrt{|{\bf P}|+deg{\bf G}^2}$.\\

{\bf Remark 2.1} In Construction A, though the coherence is slightly worse than that in \cite{DeVore,LG}, it gives us real valued measurement matrix, while only binary measurement matrices were given in \cite{DeVore,LG}.\\

\subsection{Construction B}

Let ${\bf P}_{{\bf F}_q}^2$ be the projective plane over the finite field ${\bf F}_q$,  ${\bf B}_r$ be the set of all non-singular plane curves in ${\bf P}_{{\bf F}_q}^2$ defined over ${\bf F}_q$ with their degrees equal to $r$. Then ${\bf B}_r$ is an open dense algebraic subset in the dimension $\frac{1}{2}(r+1)(r+2)-1=\frac{1}{2}r(r+3)$ projective space (Bertini's theorem, page 179, [12]). We denote the number of ${\bf F}_q$ rational points of this set as $|{\bf B}_r|$. For any non-singular plane curve $f(x,y,z)=0$ defined over ${\bf F}_q$,  the length $q^2+q+1$ vector ${\bf v(f)}_h$, where $h$ takes over all ${\bf F}_q$ rational points of the projective plane, is defined as follows. ${\bf v(f)}_h$ is zero if $h$ is not on the curve $f(x,y,z)=0$ and ${\bf v(f)}_h$ is $1$  if $h$ is on the curve $f(x,y,z)=0$. Thus the Euclid norm of this vector is exactly the number of the ${\bf F}_q$ points of this curve $f(x,y,z)=0$. Then  $||{\bf v(f)}_h|| \geq q+1-2g\sqrt{q}$ ([7, 17]) , where $g=\frac{1}{2}(r-1)(r-2)$ is the genus of this curve. On the other hand from Bezout's theorem, the inner product $|<{\bf v(f_1)}_h, {\bf v(f_2)}_h>| \leq r^2$. We have a $(q^2+q+1) \times |{\bf B}_r|$ binary matrix ${\bf \Phi}_r$. The coherence of this binary matrix $\mu_{{\bf \Phi}_r} \leq \frac{r^2}{q+1-(r-1)(r-2)\sqrt{q}}$.\\

If we take $r=2$ we have the following result.\\

{\bf Proposition 2.1.} {\em For any prime power $q$, we have a $(q^2+q+1) \times(q^5-2q^4)$ binary matrix ${\bf \Phi}_2$ with the coherence $\mu_{{\bf \Phi}_2} \leq \frac{4}{q+1}$.}\\

{\bf Proof.} We need to prove $|{\bf B}_2| \geq q^5-q^4-2q^3$. It is well-known that a plane quadrics is singular if and only if the corresponding matrix is not full rank. In the case the matrix of a plane quadrics is a symmetric $3 \times 3$ matrix. These singular plane quadrics are on a degree $3$ hypersurface in ${\bf P}_{{\bf F}_q}^5$. From the  Segre-Serre-Sorensen bound (Theorem 4.4  \cite{HK} and \cite{GKZ}) , ${\bf B}_2$ is the set of ${\bf P}_{{\bf F}_q}^5$ minus a degree $3$ hypersurface, this degree $3$ hypersurface have at most $3q^4+q^3+q^2+q+1$ ${\bf F}_q$ rational points.\\

If we take $r=3$, we need a lower bound for $|{\bf B}_3|$.\\

{\bf Proposition 2.2.} {\em For any prime power $q$, we have a $(q^2+q+1) \times |{\bf B}_3|$ binary matrix ${\bf \Phi}_3$ with the coherence $\mu_{{\bf \Phi}_3} \leq \frac{9}{q+1-2\sqrt{q}}$, where $|{\bf B}_3| \geq q^9-6q^8$.}\\

{\bf Proof.} It is well-known that a cubic plane curve is singular if and only if its discriminant, which is a degree at most $7$ polynomial of coefficients, is zero (see page 48-51 of \cite{Ellip} and appendiiex A). From Segre-Serre-Sorensen bound  (Theorem 4.4  of \cite{HK}), $|{\bf B}_3| \geq q^9-6q^8$.\\

More generally from the classical resultant theory (\cite{GKZ}) it is known that the reducible or singular plane curves of degree at most $r$ in ${\bf P}_{{\bf F}_q}^{\frac{1}{2}r(r+3)}$ are in a degree at most $r^4$ hypersurface. Thus we have $|{\bf B}_r| \geq q^{\frac{1}{2}r(r+3)}-(r^4-1)q^{\frac{1}{2}r(r+3)-1}$.\\

{\bf Proposition 2.3.} {\em We have a $(q^2+q+1) \times (q^{\frac{1}{2}r(r+3)}-(r^4-1)q^{\frac{1}{2}r(r+3)-1})$ matrix ${\bf \Phi}_d$ with its coherence at most $\frac{r^2}{q+1-(r-1)(r-2)\sqrt{q}}$.}\\

This gives us a lot of matrices with small  coherences when $r$ is relatively small compared to $q$. However this construction is worse than DeVore construction \cite{DeVore}.\\

We generalize the above construction to the case that ${\bf P}_{{\bf F}_q}^2$ is replaced by a non-singular algebraic projective surface ${\bf X}$ defined over ${\bf F}_q$. The set of all ${\bf F}_q$ rational points of this surface is denoted by ${\bf X(F_q)}$. For a very ample divisor ${\bf D}$ on ${\bf X}$, we need to use the linear system $Linear({\bf D})$ which consists of all divisors on ${\bf X}$ linearly equivalent to ${\bf D}$ (\cite{Hartshorne}, Chapter V). We denote ${\bf B(D)}$ the algebraic set of all divisors in the linear system $Linear({\bf D})$ which are non-singular curves. This is an open dense algebraic subset in the projective space of the dimension $dim(Linear({\bf D}))$ (Bertini's theorem, page 179, \cite{Hartshorne}). In many cases this dimension can be computed from the Riemann-Roch theorem of algebraic surfaces (\cite{Hartshorne}, page 362). From the adjunction formula the genus $g$ of these non-singular curves in this linear system $Linear({\bf D})$ satisfies $2g-2={\bf D}.({\bf D}+{\bf K_X})$. The number of intersection of two such non-singular curves is at most ${\bf D}.{\bf D}$. \\

For any non-singular curve $f$ in the linear system $Linear({\bf D})$, the length $|{\bf X(F_q)}|$ vector ${\bf v(f)}_h$, where $h$ takes over all ${\bf F}_q$ rational points on the surface ${\bf X}$, is defined as follows. ${\bf v(f)}_h$ is zero if $h$ is not on the curve $f$ and ${\bf v(f)}_h$ is $1$ if $h$ is on the curve $f$. Thus the Euclid norm of this vector is exactly the number of the ${\bf F}_q$ points of this curve $f$. Thus we have $||{\bf v(f)}_h|| \geq q+1-2g\sqrt{q}$, where $2g-2={\bf D}.({\bf D}+{\bf K_X})$ is the genus of these curves (\cite{Hartshorne} page 361). \\

{\bf Theorem 2.3.} {\em We have a $|{\bf X(F_q)}| \times |{\bf B(D)}|$ matrix with its coherence at most $\frac{{\bf D}.{\bf D}}{q+1-2g\sqrt{q}}$, where $g=\frac{1}{2}({\bf D}.({\bf D}+{\bf K_X}))+1$.}\\

In Theorem 2.3  ${\bf X}$ is taken as the Fermat surface in ${\bf P}_{{\bf F_{q^2}}}^3$ defined by $X_0^{q+1}+X_1^{q+1}+X_2^{q+1}+X_3^{q+1}=0$ (\cite{Hansen}). Then $|{\bf X(F_{q^2}})|=(q^3+1)(q^2+1)$ (\cite{Hansen}). Simply we take ${\bf D}$ as the hyperplane divisor, ${\bf D}.{\bf D}=(q+1)$. However we know that all hyperplane sections of the Fermat surface have at least $(q-1)^2(q+1)$ rational points (see Lemma 3.1 in section 3), and ${\bf B(D)}$ can be all these hyperplanes. Thus $|{\bf B(D)}|$ is $q^6+q^4+q^2+1$. We have a $(q^3+1)(q^2+1) \times (q^6+q^4+q^2+1)$ matrix with the coherence at most $\frac{q+1}{(q-1)^2(q+1)}=\frac{1}{(q-1)^2}$.\\

\subsection{Construction C}

Let ${\bf Y}$ be a non-singular algebraic projective manifold defined over ${\bf F}_q$. The set of all ${\bf F}_q$ rational points of this manifold is denoted by ${\bf Y(F_q)}$. For an effective divisor ${\bf D}$ on ${\bf Y}$, we will use  the function space $L({\bf D})$ which consists of all rational functions on ${\bf Y}$ with poles at most $-{\bf D}$ (\cite{Hartshorne}).  In many cases the dimension of this function space can be computed from the Riemann-Roch theorem (\cite{Hartshorne}). For any rational function $f \in L({\bf D})$,  the length $q \cdot |{\bf Y(F_q)}-{\bf D}|$ vector ${\bf v(f)}_h$, where $h=(a,b)$, $ b \in {\bf Y(F_q)}-{\bf D}$ and $a \in {\bf F}_q$, is defined as follows. ${\bf v(f)}_h$ is zero if $h=(a,b)$ satisfy $f(b) \neq a$,  and ${\bf v(f)}_h$ is $1$  if $h=(a,b)$ satisfy $f(b)=a$. Thus the Euclid norm of this vector is exactly $|{\bf Y(F_q)}-{\bf D}|$. The inner product satisfies $|<{\bf v(f_1)}_h, {\bf v(f_2)}_h>|$ is at most the number of zero points in ${\bf Y(F_q)}$ of the function $f_1-f_2$. That is, the absolute value of this inner product is smaller or equal to the maximal possible number of ${\bf F}_q$ rational points of members of the linear system $Linear({\bf D})$. We denote this number by $N({\bf D})$. \\

{\bf Theorem 2.4.} {\em We have a $q \cdot |{\bf Y(F_q)}-{\bf D}| \times q^{dim(L({\bf D}))}$ matrix with the coherence at most $\frac{N({\bf D})}{|{\bf Y(F_q)}-{\bf D}|}$.}\\

For example ${\bf Y}={\bf P}_{{\bf F}_q}^n$ and ${\bf D}=r{\bf H}$, where ${\bf H}$ is the hyperplane divisor of the projective space. Then $|{\bf Y(F_q)}-{\bf D}|=q^n$, $dimL( r{\bf H})=\displaystyle{n+r \choose r}$, $N({\bf D}) \leq rq^{n-1}+q^{n-2}+q^{n-3}+\cdots+q+1$ (Segre-Serre-S\/orensen bound, \cite{HK}). We have $q^{n+1} \times q^{\displaystyle{n+r \choose r}}$ matrix with the coherence at most $\frac{rq^{n-1}+q^{n-2}+q^{n-3}+\cdots+q+1}{q^n}$. When $n$ and $r$ are quite close to $q$ this example is better than the outputs of the DeVore construction. \\

\section{Examples}

{\bf 3.1. Fermat surface.} In this example we need the following Lemma.\\

{\bf Lemma 3.1.} {\em Let $t< q+1$ be a positive integer satisfying $gcd(q^2-1, t)=1$. For any degree $t$ hypersurface ${\bf Y}_t$ in ${\bf P}_{{\bf F}_{q^2}}^3$, there are at least $(q-1)^2(q+1)$ ${\bf F}_{q^2}$ rational points in the intersection of ${\bf Y}_t$ and the Fermat surface $X_0^{q+1}+X_1^{q+1}+X_2^{q+1}+X_3^{q+1}=0$.}\\

{\bf Proof.} For any degree $t$ hypersurface defined by a homogeneous polynomial  $f(X_0,X_1,X_2,X_3)=0$, when we fix $(q-1)(q+1)$ possibilities of $X_0$ and $X_1$ as $X_0 \theta^{i(q-1)}$ and $X_1 \theta^{i(q-1)}$, $i=0,...,q$ (here $\theta$ is a primitive element of the multiplicative group ${\bf F}_{q^2}^*$.), we note that we get two equations $X_2^{q+1}+X_3^{q+1}=a$ and $a_tX_2^t+a_{t-1}X_2^{t-1}X_3+\cdots+a_1X_2X_3^{t-1}+a_0X_3^t=d$, where $a \in {\bf F}_q$ and $d \in {\bf F}_{q^2}$. Then $a_t (\frac{X_2}{X_3})^t+\cdots+a_1\frac{X_2}{X_3}+a_0=\frac{d}{X_3^t}$. When $X_3$ is changed to $X_3\theta^{i(q-1)}$, $i=0,...,q$, the first equation from Fermat surface is satisfied.  There are $q+1$ possibilities of $\frac{d}{X_3^t}$ since $gcd(q^2-1,t)=1$. Thus if $t< q+1$ there are at least $\frac{q^2}{t} \cdot (q+1)$ images and  there are at least one solution for the second equation. On the other hand from the homogeneousity we have at least $\frac{(q-1)^2(q+1)^2(q-1)}{q^-1}=(q-1)^2(q+1)$ such solutions.\\

From Construction B, if the conditions $t <q$ and $gcd(t,q^2-1)=1$ are satisfied we have $(q^2+1)(q^3+1) \times O(q^{\frac{1}{3}(t+1)(t+2)(t+3)-1})$ matrix with the coherence $\frac{t^2(q+1)}{(q-1)^2(q+1)}=\frac{t^2}{(q-1)^2}$ when ${\bf D}=t{\bf H}$ is taken, here ${\bf H}$ is the hyperplane section of the Fermat surface (\cite{Hansen}).  Here we should note that the linear system $Linear(t{\bf H})$ on ${\bf X}$ is the same as the linear system of degree $t$ homogeneous polynomials on ${\bf P}_{{\bf F}_{q^2}}^3$ (\cite{Hansen}. When $c\sqrt{q} \leq t < q+1$, $c$ is a suitable positive constant, and $gcd(t,q^2-1)=1$, the matrices from this construction have smaller coherences than that of matrices from the DeVore construction \cite{DeVore}.\\

{\bf 3.2. Projective spaces.} From construction C, we take ${\bf P}_{{\bf F}_p}^9$  then we have $p^{10} \times p^{\displaystyle{9+r \choose r}}$ matrix with the coherence at most $\frac{r}{p}+\frac{1+\frac{1}{p}+\frac{1}{p^2}+\cdots+}{p^2} \leq \frac{r}{p}+\frac{2}{p^2}$ when $p$ tends to the infinity. On the other hand if we want to get this matrix from Hermitian curve using the construction in \cite{LG} (page 5040 of \cite{LG}) , set $s=\frac{\displaystyle{9+r \choose r}}{4}+g-1$, $g=\frac{p^4-p^2}{2}$. The coherence of this matrix from the construction is $\frac{s}{p^6}$. When $c$ is a small positive real constant satisfying $s=\frac{\displaystyle{9+r \choose r}}{4}+g-1 < p^6$. We have  $\displaystyle{9+r \choose 9} \approx r^9$. Set $r=cp$ where $c$ is a positive constant satisfying $p^{-2/3}<c<p^{-1/3}$ (then $p^3<s < p^6$, Hermitian curve RIP matrices have smaller coherences than DeVore construction RIP matrices). The ratio $\frac{\mu_{Hermitian}}{\mu_{constructionC}}=\frac{c^8p^3}{8} \approx \frac{p^{1/3}}{8}$. When $p$ is a prime number and tends to the infinity, this ratio tends to the infinity. This illustrate that our construction C is much better than the direct curve-based generalization \cite{LG} of the DeVore construction in \cite{DeVore}.\\

{\bf 3.3. Ruled Surface.} We take ${\bf X}={\bf P}_{{\bf F}_q}^1 \times {\bf P}_{{\bf F}_q}^1$ in construction C. The counting of rational points of members in a linear system $Linear({\bf D})$ on a surface is treated as follows in \cite{Hansen}. If the ${\bf F}_q$ rational points of ${\bf X}$ are distributed on several ${\bf F}_q$ rational curves ${\bf C}_1,....,{\bf C}_h$ in ${\bf X}$. Then we count how many of curves can be in a member of this linear system and count the intersection numbers of the divisor ${\bf D}$ with these curves ${\bf C}_i$, $i=1,...,h$. In this case the set of ${\bf F}_q$ rational points on ${\bf P}_{{\bf F}_q}^1 \times {\bf P}_{{\bf F}_q}^1$ is naturally the disjoint union of $(q+1)$ sets of ${\bf F}_q$ rational points on curves $p_i \times {\bf P}_{{\bf F}_q}^1$, where $p_i$, $i=1,...,q+1$ are $(q+1)$ rational points of ${\bf P}_{{\bf F}_q}^1$. Thus if we take the divisor ${\bf D}$ of type $(d_1,d_2)$, that is, polynomials $f(x,y,z,w)$ which are homogeneous in $x,y$ with degree $d_1$ and is homogeneous in $z,w$ with degree $d_2$, we get a linear system with dimension $(d_1+1)(d_2+1)$. If $d_1+d_2 <q+1$, there are at most $-d_1d_2+d_1(q+1)+d_2(q+1)$ rational points on any member of this linear system. (\cite{Hansen}). We have a $q^3 \times q^{(d_1+1)(d_2+1)}$ RIP matrix with the coherence at most $\frac{-d_1d_2+(d_1+d_2)(q+1)}{q^2}$. \\

Set $q=p^{\frac{10}{3}}$, we have a $p^{10} \times p^{\frac{10}{3}(d_1+1)(d_2+1)}$ matrix with the coherence at most $\frac{-d_1d_2+(d_1+d_2)(p^{10/3}+1)}{p^{20/3}}$. We set $d_1$ and $d_2$  in the range $p^{8/3}< d_1, d_2 <p^3$ (and then $d_1d_2 \approx s \leq p^6$ as required in \cite{LG}. If $d_1=d_2=cp^{\frac{8}{3}}$ for suitable prime number $p$ and suitable positive number $c$ satisfying $0< c < p^{\frac{1}{3}}$. For the RIP $p^{10} \times p^{(d_1+1)(d_2+1)}$ matrix from Construction C. the coherence is $\frac{(d_1+d_2)(p^{\frac{10}{3}}+1)-d_1d_2}{p^{\frac{20}{3}}} \approx \frac{2c}{p^{2/3}}-\frac{1}{p^{4/3}}$.\\

 If we want to get this matrix from the construction \cite{LG} (page 5040) from Hermitian curve, the degree of the divisor $s=\frac{5(d_1+1)(d_2+1)}{6}+\frac{p^4-p^2}{2}-1 \approx \frac{5c^2p^{16/3}}{6}+\frac{p^4-p^2}{2}$. Then the upper bound of the coherence from Hermitian curve construction in \cite{LG} is $\frac{s}{p^6} \approx \frac{5c^2}{6p^{2/3}}+\frac{1}{p^2}-\frac{1}{p^4}$. Then the ratio of two coherence is $\frac{\mu_{constructionC}}{\mu_{Hermitian}} \approx 5/12c$. If $\frac{12}{5}<c<p^{1/3}$. Then the coherence of our construction C is much better. When $p$ tends to the infinity, the ratio tends to $0$.\\

{\bf 3.4. Deligne-Lusztig surface.} Deligen-Lusztig varieties were introduced in \cite{DL} and had been found very useful in coding theory (\cite{Hansen}) since there are many rational points on them. In this section we use the Deligne-Lusztig surface ${\bf X}$ of the type $A_4^2$ defined over the finite field ${\bf F}_{q^2}$ as in \cite{Hansen} section 4 pages 542-545 to construct RIP matrices. There are $(q^5+1)(q^3+1)(q^2+1)$  rational points on ${\bf X}$. The divisor ${\bf L}$ as in \cite{Hansen} section 4 is used. It is the pull-back of the $t{\bf H}$ where ${\bf H}$ is the hyperplane section on ${\bf P}_{{\bf F}_{q^2}}^4$ and $t$ satisfying $t <q^3+1$ as required in section 4 of \cite{Hansen}. Then in the linear system $Linear({\bf L})$ there are at most $t(
(q^5+1)(q^3+1)+(q+1)(q^3+1)(q^2+1-t))$ rational points. The dimension of this linear system is $\displaystyle{4+t \choose 4}-\displaystyle{4+t-(q+1) \choose 4} \approx cqt^3$ where $c$ is a positive constant. From construction C we have $q^2(q^3+1)(q^7+q^5-q^3+1)  \times q^{cqt^3}$ RIP matrix with the coherence $\frac{t((q^5+1)(q^3+1)+(q^3+1)(q+1)(q^2+1-t))}{(q^3+1)(q^7+q^5-q^3+1)} \approx \frac{t}{q^2}$. Set $q=p^{5/6}$, when $t$ is in the range $p^{21/12} <t <p^{5/2}$, we have $\frac{\mu_{ConstructionC}}{\mu_{Hermitian}}$ tends to $0$ when $p$ tends to the infinity ($\mu_{Hermitian}$ as in page 5040  of \cite{LG}).\\

{\bf 3.5. Toric surfaces.}  Toric varieties are typtical geometric objects in algebraic geometry and complex differential geometry (\cite{Fulton} and played an interesting role in algebraic-geometric coding theory (\cite{JHansen}). In this section we give some RIP matrices from toric surfacses and show that their coherence are quite small.\\

Let ${\bf Z}^2 \subset {\bf R}^2$ be the set of all integral points . We denote $\theta$ a primitive element of the finite field ${\bf F}_q$. For any integral point ${\bf m}=(m_1, m_2) \in {\bf Z}^2$ we have a function $e({\bf m}): {\bf F}_q^{*} \times {\bf F}_q^{*} \rightarrow {\bf F}_q$ defined as $e({\bf m})(\theta^i,\theta^j)=\theta^{m_1i+m_2j}$ for $i=0,1,...,q-1$ and $j=0,1,...,q-1$. Let $\Delta \subset {\bf R}^2$ be a convex polyhedron with vertices in ${\bf Z}^2$ and $L(\Delta)$ be the function space over ${\bf F}_q$ spanned by these functions $e({\bf m})$ where ${\bf m}$ takes over all integral points in $\Delta$. In the following cases of convex polyhedrons these functions are linearly independent from the result in \cite{JHansen}.\\

For each function $f \in L(\Delta)$ we have a length $q \times (q-1)^2$ vector ${\bf v}(f)=(f_{(a,b)})$ where $(a,b) \in {\bf F}_q^{*} \times {\bf F}_q^{*} \times {\bf F}_q$ defined as follows. $f_{(a,b)}=0$ if $f(a) \neq b$ and $f_{(a,b)}=1$ if $f(a)=b$. Then we have $q^{dim(L(\Delta))}$ such vectors and a $q(q-1)^2 \times q^{dim(L(\Delta))}$ matrix ${\bf \Phi}_{\Delta}$. The following cases as in  the main results Theorem 1, 2, 3 of \cite{JHansen} are considered.\\
1) $\Delta$ is the convex polytope with the vertices $(0,0), (d,0), (0,d)$ where $d$ is a positive integer satisfying $d <q-1$;\\
2) $\Delta$ is the convex polytope with the vertices $(0,0), (d,0), (d,e+rd), (0,e)$ where $d, r, e$ are positive integers satisfying $d<q-1$, $e<q-1$ and $e+rd<q-1$;\\
3) $\Delta$ is the convex polytope with the vertices $(0,0), (d,0), (0,2d)$ where $d$ is a positive integer satisfying $2d <q-1$;\\

 We have the following result from the main results Theorem 1, 2, 3 of \cite{JHansen}.\\

{\bf Proposition 3.1.} {\em In the above cases the matrix ${\bf \Phi}_{\Delta}$ is a RIP matrix whose coherence satisfying the following\\
1) $\mu({\bf \Phi}_{\Delta}) \leq \frac{d}{q-1}$ in the case 1);\\
2) $\mu({\bf \Phi}_{\Delta}) \leq \frac{min\{(d+e)(q-1)-de, (e+rd)(q-1)\}}{(q-1)^2}$ in the case 2);\\
3) $\mu({\bf \Phi}_{\Delta}) \leq \frac{2d}{q-1}$ in the case 4);\\
The size parameter $dim(L(\Delta))$ is the number of the integral points in $\Delta$. That is, \\
1) $dim (L(\Delta))=\frac{(d+1)(d+2)}{2}$ in case 1);\\
2) $dim (L(\Delta))=(d+1)(e+1)+\frac{rd(d+1)}{2}$ in case 2);\\
3) $dim (L(\Delta))=d^2+2d+1$ in case 3).}\\

Comparing with Exampel 3.3 for  these RIP matrices from toric surfacses the coherence is smaller in some range of paramters. Moreover the coherence is better than that of RIP matrices from DeVore construction \cite{DeVore} and its direct generalization \cite{LG} in some range of parameters.\\

\section{$\pm1$-Randomized RIP matrices from algebraic geometry}

We observe that if the $1$'s coordinates in DeVore and our construction are changed to $-1$ randomly, the conclusion about the coherence is still true.\\

{\bf Proposition 4.1.} {\em We  get  RIP matrices with $\mu_{randomized} \leq \frac{r}{p}$ in DeVore construction and  $\mu_{randomized}  \leq \frac{N({\bf D})}{|{\bf Y(F_q)}-{\bf D}|}$ in Construction C if the coordinates $1$'s are changed to $\pm1$'s randomly.}\\

{\bf Proof.} For any two columns ${\bf v(f)}$ and ${\bf v(g)}$ in the constructed matrix, the  number of the common non-zero coordinates  is at most $r$ in DeVore construction and $N({\bf D})$ in our  construction C. The Euclid norms of these columns are not changed. The conclusion follows directly.\\

The following result is about the mathematical expectation of coherence after this $\pm1$-randomization of DeVore and our C constructions.\\

{\bf Theorem 4.1.} {\em The probability that the coherence of the $\pm1$-randomized DeVore construction satisfying  $\mu_{randomized} \leq \frac{1}{p}$ is at least $(\frac{1}{2})^r$ and the mathematical expectation  of the coherence is
$$
\begin{array}{ccc}
{\bf E(\mu_{randomized})} \leq \frac{r-4\Sigma_{w=0}^{[\frac{r}{2}]} w \cdot \displaystyle{r \choose w} \cdot (\frac{1}{2})^r}{p}
\end{array}
$$
. Similarly the probability that the coherence of the $\pm1$-randomized  Construction C satisfying $\mu_{randomized} \leq \frac{1}{|\bf Y(F_q)-{\bf D}|}$ is at least $(\frac{1}{2})^{N({\bf D})}$ and the mathematical expectation of the coherence is

$$
\begin{array}{ccc}
{\bf E(\mu_{randomized})} \leq  \frac{ N({\bf D})-4\Sigma_{w=0}^{[\frac{N({\bf D}}{2}]} w \cdot \displaystyle{N({\bf D}) \choose w} \cdot (\frac{1}{2})^{N({\bf D})}}{|{\bf Y(F_q)}-{\bf D}|}
\end{array}
$$.}

{\bf Proof.}  Suppose there are $L$   common non-zero coordinates  for two columns of the constructed matrix ${\bf v(f)}$ and ${\bf v(g)}$. Then the possibility that the two columns have the same $1$ or $-1$ at one common position is $\frac{1}{2}$, and the possibility that the two columns have the different $1$ and $-1$ at this position is $\frac{1}{2}$. Thus the possibility that the inner product $<{\bf v(f)}, {\bf v(g)}>= \pm( L-2w)$ is $\displaystyle{L \choose w} (\frac{1}{2})^L$. The mathematical expectation of the inner product ${\bf E}(<{\bf v(f)}, {\bf v(g)}>)=\Sigma_{w=0}^L |L-2w| \displaystyle {L \choose w} \frac{1}{2^L}=L-4\Sigma_{w=0}^{[\frac{L}{2}]} w \displaystyle{L \choose w} (\frac{1}{2})^L$ when $L$ is odd and $L-(\frac{1}{2})^L \displaystyle{L \choose \frac{L}{2}}-4\Sigma_{w=0}^{\frac{L}{2}-1} w \displaystyle{L \choose w} (\frac{1}{2})^L$ when $L$ is even. The conclusion follows directly.\\

In random construction of RIP matrices, we only can get RIP matrices with high probabilty. In our $\pm1$-randomized RIP matrices from algebraic geometry, the constructed matrices satisfy the RIP property deterministically and the upper bound of coherence can be derived from counting of rational points. Moreover if $\pm1$ coordinates are randomized and the sizes and sparsity orders are large, the coherences are much smaller in the sense of average. Thus in practical signal processing it is more suitable to use $\pm1$-randomized RIP matrices from algebraic geometry.\\

\section{Strong coherence property}

In this section we first prove the following two results.\\

{\bf Theorem 5.1.} {\em The RIP matrices from DeVore construction and its direct AG generalization do not satisfy the strong coherence property.}\\

{\bf Proof.} We need to calculate the sum $\Sigma_{deg(f) \leq r-1} \frac{v_f}{||v_f||}$. Since $||v_f||^2=p$ for each polynomial $f \in {\bf F}_p[x]$, $\Sigma_{deg(f) \leq r-1} \frac{v_f}{||v_f||}=\frac{1}{p^2} \Sigma_{deg(f) \leq r-1} {v_f}$. On the other hand for each point $(a, b) \in {\bf F}_p \times {\bf F}_p$, there are exactly $p^{r-1}$ polynomials $f$'s in ${\bf F}_p[x]$ with degrees $deg(f) \leq r-1$ satisfying $f(a)=b$. We have $\Sigma_{deg(f) \leq r-1} {v_f}=p^{r-1} {\bf 1}$, where ${\bf 1}$ is a length $p^2$ vector with all coordinates $1$. Thus $\omega_{\bf \Phi}=\frac{p^{r-1}-1}{p^r-1}$.  In the DeVore construction  $\mu_{{\bf \Phi}}=\frac{r}{p}$. The conclusion follows from the condition $ r <p$.\\

For the direct algebraic-geometric generalization in \cite{LG}, similarly we have $||v_f||^2=|{\bf P}|$. For each $(a,P_i)$, the coordinate of $\Sigma_{f \in L({\bf G})} \frac{v_f}{||v_f||}$ at $(a,P_i)$ is $q^{dimL(G)-1}$ if not all functions in $L(G)$ are zero at $P_i$. Thus $\Sigma_{f \in L({\bf G})} \frac{v_f}{||v_f||}=q^{dimL({\bf G})-1}{\bf 1}$. Then $\omega_{{\bf \Phi}}=\frac{q^{dim(L({\bf G})-1}-1}{q^{dimL({\bf G})}-1}$. The conclusion follows directly from the fact $dimL({\bf G})=deg{\bf G}-g+1$.\\

{\bf Theorem 5.2.} {\em The constructed measurement matrices in Construction C does not satisfy the strong coherence property.}\\

{\bf Proof.} Since $||{\bf v(f)}||^2=|{\bf Y(F_q)}|-{\bf D}|$. On the other hand for each point $(a,b)$ in ${\bf F}_q \times ({\bf Y(F_q)}-{\bf D})$, there are exactly $q^{dim(L({\bf D})-1)}$ rational functions $f$ satisfying $f(b)=a$. We have $\omega_{{\bf \Phi}}=\frac{q^{dim(L({\bf D})-1}-1}{q^{dimL({\bf D})}-1}$. Thus it has no strong coherence property.\\

In the construction of DeVore \cite{DeVore} and our construction C, the coordinates of $1$ can be changed to $\pm1$ randomly and the conclusion about the coherence is valid. We give some results of this $\pm1$-randomized AG construction in the previous section. If the $\pm1$ could be arranged properly, this would give us the RIP matrices with small coherence and satisfying the strong coherence property.\\

{\bf Modified AG-RIP matrices in compressed sensing.}\\

An even number  of $\pm1$'s are said balanced if the numbers of the $1$ and $-1$ are equal and an odd number of $\pm1$'s are said balanced if the difference of the numbers of $1$ and $-1$ is $\pm1$.\\

{\bf A. Balanced $\pm1$ in each ${\bf v(f)}$.}\\

We divide the set ${\bf Y(F_q)}-{\bf D}$ to two balanced parts, one part contains $[\frac{|{\bf Y(F_q)}-{\bf D}|}{2}]$ points, we color these points as red. The another part contains $|{\bf Y(F_q)}-{\bf D}|-[\frac{|{\bf Y(F_q)}-{\bf D}|}{2}]$ points, we color these points as blue. \\

{\bf B. Balanced $\pm1$ at each point $(a, b) \in {\bf F}_q \times ({\bf Y(F_q)}-{\bf D})$.}\\

We fix a base of the function space $L({\bf D})$ as $f_1,...,f_T$ where $T=dim(L({\bf D}))$. For each function $f \in L({\bf D})$, set $f=a_1f_1+\cdots+a_Tf_T$.  Suppose $q=p^s$. If $p$ is an odd prime, let $Tr: {\bf F}_q \rightarrow {\bf F}_p=\{0,...,p-1\}$ be the trace function. The parity is determined depending on the parity of the element in $\{0,1,...,p-1\}$. We will use the parity of the element $Tr(a_1+\cdots+a_T)$ to balance the $\pm1$'s at each point $(a,b)$.\\
In the case $p=2$ we use the parity of the Hamming weight $wt((Tr(a_2),...,Tr(a_T))$, that is, the number of $1$'s in $(Tr(a_2),...,Tr(a_T))$. Here without loss of generality we assume $f_1(b) \neq 0$.\\

{\bf Assignment of $\pm1$ for each ${\bf v(f)}$.}\\

Suppose $p$ is an odd prime. For each column ${\bf v(f)}$ and each $(a, b) \in {\bf F}_q \times ({\bf Y(F_q)}-{\bf D})$ satisfying $f(b)=a$, we set $(-1)^{\lambda+parity(Tr(a_1+\cdots+a_T)}$ if $p$ is odd and $(-1)^{\lambda+parity(wt(Tr(a_2),...,Tr(a_T)))}$ (we assume that $f_1(b) \neq 0$) if $p=2$.
  Here $\lambda=1$ if $b$ is red and $\lambda=-1$ if $b$ is blue. ${\bf v(f)}=0$ at the point $(a,b) \in {\bf F_q} \times ({\bf Y(F_q)}-{\bf D})$ if $f(b) \neq a$. Then in each column ${\bf v(f)}$, it is obvious $\pm1$'s are balanced from A. On the other hand we have the following Lemma.\\

{\bf Lemma 5.1.} {\em For each point $(a, b) \in {\bf F}_q \times ({\bf Y(F_q)}-{\bf D})$ and $q^{T-1}$ rational functions $f \in L({\bf D})$ satisfying $f(b)=a$, the coordinates of  $p^{(T-1)s-1}$ columns at $(a, b)$ is $1$, the $\pm1$'s of other $p^{(T-1)s}-p^{(T-1)s-1}$ columns are balanced.}\\

{\bf Proof.} Suppose $a_1^0f_1(b)+\cdots+a_T^0f_T(b)=a$, then the set of all vectors $(a_1-a_1^0,...,a_T-a_T^0)$ satisfying $a_1f_1(b)+\cdots+a_Tf_T(b)=a$  is a dimension $T-1$ subspace ${\bf F}_q^{T-1}$ of $L({\bf D})={\bf F}_q^T$. Therefore the set of all $(a_1,...,a_T) \in {\bf F}_q^{T-1}$ satisfying  $Tr(a_1-a_1^0+\cdots+a_T-a_T^0)=0$ is a dimension $s(T-1)-1$ space over ${\bf F}_p$. There are $p^{(T-1)s-1}$ columns such that the coordinates of these columns at $(a,b)$ is $1$. In the case $p$ is an odd prime, set $\phi: {\bf F}_p=\{0,...,p-1\} \rightarrow {\bf F}_p$, $\phi(x)=-x$. It is clear that $\phi$ changes the parity.  In this case the 2nd conclusion follows from the linearity of $Tr$ over ${\bf F}_p$. When $p=2$, since $f_1(b) \neq 0$, the variables $a_2,..., a_T$ are free and the  conclusion follows directly.\\

{\bf Theorem 5.3.} {\em  The RIP matrices from modified DeVore construction satisfy the strong coherence property if $r(r-1) \leq \frac{q}{160logq}$. If  a) $N({\bf D}) > \frac{\sqrt{|Y(F_q)-{\bf D}|}}{p\sqrt{q}}$ and b) $dim(L({\bf D})) \leq \frac{|{\bf Y(F_q)}-{\bf D}|}{160logq}$ are satisfied, the RIP matrices from the modified construction C has the strong coherence property. When $p$ is fixed and $q=p^s$ is sufficiently large, this condition a) is valid.}\\

{\bf Proof.} From Lemma 5.1 we have $\Sigma_{j=1, j\neq i}^{dim(L({\bf D})}  \frac{<{\bf \phi_j}, {\bf \phi}>}{||{\bf \phi_j}|| \cdot ||{\bf \phi_i}||} \leq p^{(T-1)s-1}-1$. Thus the average coherence is at most $\frac{p^{(T-1)s-1}-1}{p^{sT}-1} \leq \frac{p^{(T-1)s-1}}{p^{Ts}}=\frac{1}{p^{s+1}}$. In the modified DeVore construction it is clear that  the coherence is exactly $\frac{r-1}{p}$. The 1st conclusion follows directly. In the case of construction C, it is direct from our assignment of $\pm1$'s that the coherence is exactly $\frac{N({\bf D})}{|{\bf Y(F_q}-{\bf D}|}$. The 2nd conclusion follows from Weil-Deligne theorem \cite{Deligne} or Lachaud bound \cite{GL} directly.\\

In the case ${\bf Y}$ is projective space ${\bf P}_{F_q}^n$ and ${\bf D}=r{\bf H}$ where ${\bf H}$ is the hyperplane section. Then the condition a) in Theorem 5.3  is $N(r{\bf H}) \geq \frac{q^{(n-1)/2}}{p}$. This condition is satisfied automatically. The condition b) in Theorem 5.3 is $\displaystyle{n+r  \choose r} \leq \frac{q^n}{160logq}$. This condition can be satisfied when $r/q$ is relatively small. In Example 3.3 the condition b) in Theorem 5.3 is $(d_1+1)(d_2+1) \leq \frac{q^2}{160log q}$. The condition a) is $(d_1+d_2)(q+1)-d_1d_2  \geq \frac{\sqrt{q}}{p}$. The condition a) and b) are satisfied if $d_1$ and $d_2$ are $O(q^t)$ where $t<1$. In Example 3.4 Deligne-Lusztig surface case if $t \leq q^3$, the condition b) is satisfied and the condition a) is satisfied automatically. For RIP matrices from toric surfaces in 3.5 the condition a) is satisfied automatically and the condition b) is not difficult to be satisfied.\\

From above examples we can see that the conditions a) and b) in Theorem 5.3 are not difficult to be satisfied for most  algebraic varieties and very ample divisors. Thus from our modified algebraic geometric construction of RIP matrices, there are many candidates satisfying the small coherence property and strong coherence property for the practical use.\\

\section{Asymptotic bound}

{\bf Theorem 6.1.} {\em From our construction C, we can construct  $n \times N$ RIP matrices from algebraic curves over ${\bf F}_{q^2}$ satisfying the aysmptotic bound  $\mu=O((\frac{logN}{nlog(n/logN)})^{1/3})$ in the  parameter range $logN \leq n \leq (logN)^4$.}\\

{\bf Proof.} From the Drinfeld-Vladut bound (\cite{TV}), there is a family of projective curves ${\bf X}_h$ with genus $g_h$ and $N_n$ rational points over the fixed finite field ${\bf F}_{q^2}$ satisfying $lim\frac{N_h}{g_h}=q$. On each such curve ${\bf X}_h$ we take a rational divisor ${\bf D}_h$ of degree $t_hg_h$ satisfying $2<t_h<q$. We consider the $w$ dimensional algebraic variety ${\bf X}_h \times \cdots \times {\bf X}_h$ ($w$ copy product) and the divisor ${\bf D}=\Sigma {\bf X}_h \times \cdots \times {\bf D}_h  \times \cdots \times {\bf X}_h$. Then $dim(L({\bf D})) \geq ((t_h-1)g_h)^w$. On the other hand there are at most $wt_h|{\bf X_h(F_{q^2}}|^{w-1}$ rational points on each member of the linear system $Linear({\bf D})$. Thus we have a $q^2 N_h^w \times q^{2((t_h-1)g_h)^w}$ RIP matrices with the coherence at most $\frac{wt_hg_h|{\bf X(F_{q^2}}|^{w-1}}{|{\bf X_h(F_{q^2})}|^w}=\frac{wt_hg_h}{|{\bf X_h(F_{q^2}}|} $. It is clear  when $2(t_h-1)^w \leq q^{w+2} \leq 16(t_h-1)^{4w}$ the RIP matrix size is in the range $log N \leq n \leq (logN)^4$.\\

On the other hand $(\frac{logN}{nlog(n/logN)})^{1/3} \approx  \frac{(t_hg_h)^w}{(q^{w+2}g_h^w)^{1/3}}\approx \frac{t_h^{w/3}}{q^{(w+2)/3}}$ when $g_h$ tends to the infinity. $\frac{wt_hg_h}{|{\bf X_h(F_{q^2})}|} \approx \frac{wt_h}{q}$ when $g_h$ tends to the infinity. Thus when $w, t_h, q$ are constants  the conclusion follows directly.\\

\section{Summary}

Explicit RIP measurement matrices are needed in practical application of compressed sensing to signal processing. In this paper general method of contructing explicit RIP matrices from general algebraic varieties over finite fields are presented. Many examples of RIP matrices with better performace than previous works are given. We also indicate $\pm1$-randamization of these RIP matrices from algebraic geometry are more suitable for practical compressed senseing in the presence of noisy.\\

\bibliographystyle{amsplain}

\begin{thebibliography}{10}

\bibitem{Alon} N. Alon, O. Goldreich, J. Hastad and R. Perala, Simple constructions of almost k-wise independent random variables, Random Structures Algorithms, vol.3 (1992), 289–304.
\bibitem{Amini} A. Amini and V. Montazerhodjat and F. Marvasti, Matrices with small conherence using $p$-ary block codes, IEEE Transactions on Signal Processing, Vol.60 (2012), no.1, 172-181.
\bibitem{AHSC} L. Applebaum, S. Howard, S.Searle and R. Calderbank, Chirp sensing codes: Deterministic compressed measurement for fast recovery, Applied Comptational Harmonic Analysis, Vol.26 (2009), no.2, 283-290.
\bibitem{BCJ} W.U. Bajwa, R. Calderbank, S. Jafarpour, Why Gabor frames? Two fundamental measures of coherence and their role in model selection, J. Commun.
Netw. Vol.12 (2010) 289–307.
\bibitem{BCM} W.U. Bajwa, R.Calderbank and D.G. Mixon, Two are better than one, Fundamental parameters of frame coherence, Applied and Computational Harmonic Analysis, Vol. 33 (2012), 58-78.
\bibitem{Ben} A. Ben-Aroya, and A. Ta-Shma, Constructing small-bias sets from algebraic- geometric codes,  50th Annual IEEE Symposium on Foundations of Computer Science (Atlanta, 2009), IEEE Computer Soc., Los Alamitos, Calif., 2009,191–197.
\bibitem{BDFK} J.Bourgain, S.Dilworth, K.Ford, S.Konyagin and D.Kutzarova, Explicit construction of RIP matrices and related problems, Duke Mathematical Journal, Vol.159 (2011), no.1, pages 145-185.
\bibitem{CHJ} R. Calderbank, S. Howard and S. Jafarpour, Construction of a large class of deterministic sensing matrices that satisfy a statistical isometry property, IEEE Journal of Selected Topics in Signal Processing, Vol. 4(2010), 358-374.
\bibitem{CTY} R. Calderbank, A. Thompson and Yao Xie, On block coherence of frames, Applied and Computational Harmonic Analysis, Vol.38 (2015), 50-71.
\bibitem{CRT} E.J. Cand\'es, J.Romberg and T.Tao, Robust uncertainty principles: exact signal reconstruction from highly incomplete frequency information, IEEE Transactions on Information Theory, Vol.52 (2006), no.2, pages 489-509.
\bibitem{CT} E. J. Cand\'es and T. Tao. Reflections on compressed sensing. IEEE Information Theory Society Newsletter, Dec 2008 58(4), 14--17.
\bibitem{Candes} E. J. Cand\'es. The restricted isometry property and its implications for compressed sensing. Compte Rendus de l'Academie des Sciences, Paris, Serie I, 346 (2008) 589-592.
\bibitem{Candes1} E. J. Cand\'es, Compressive sampling, Proceeding ICM 2006.
\bibitem{CLMW} E. J. Cand\'es, X. Li, Y. Ma, and J. Wright, Robust Principal Component Analysis? Journal of ACM, Vol. 58 (2009), no1., 1-37.
\bibitem{CENR} E. J. Cand\'es, Y. C. Eldar, D. Needell and P. Randall, Compressed Sensing with Coherent and Redundant Dictionaries, Applied and Computational Harmonic Analysis, Vol. 31(2011), no.1, 59-73.
\bibitem{Candes2} E. J. Cand\'es, Mathematics of sparsity (and a few other things), Proceeding ICM 2014.
\bibitem{CDD} A Cohen, W. Dahmen and R. DeVore, Compressed sensing and best $k$-term approximation, Journal of American Mathematical Society, Vol. 22 (2009), no.1, 211-231.
\bibitem{DL} P.Deligen and G. Lusztig, Representations of reductive groups over finite fields, Annals of Mathematics, (2), Vol.103 (1976), 103-161.
\bibitem{Deligne} P. Deligne, La conjecture de Weil I, Publ. IHES, vol.43 (1973), 273-307.
\bibitem{DeVore} R. DeVore, Deterministic constructions of compressed sensing matrices, Journal of Complexity, Vol.23 (2007), no.46,  918-925.
\bibitem{Donoho} D. Donoho, Compressed sensing, IEEE Transactions on Information Theory, vol.52 (2006), no.4, 1289-1306.
\bibitem{Fulton} W. Fulton, Introduction to toric varieties, Annals of Mathematics Studies, no. 131, Princeton University Press, 1993.
\bibitem{GL} S. R. Ghorpade and G. Lachaud, Etale cohomology, Lefstchetz theorems and number of points of singular varieties over finite fields, Moscow Mathematical Journal, Vol. 2 (2002), no.3, 589-631.
\bibitem{GKZ} I. M. Gelfand, M.M.Kaparanov and A.V. Zelevinsky, Discriminants, resultants and multidimensional determinants, Modern Birkhauser Classics, Birkauser, 1994, Chapter 13.
\bibitem{JHansen} J. P. Hansen, Toric surfaces and codes, techniques and examples, Coding theory, cryptography and related areas, ed. J. Bachmann et al., Springer, 2000.
\bibitem{Hansen} S. H. Hansen, Error-correcting codes over higher dimensional varieties,  Finite Fields and Their Applications,Vol. 7 (2001), 530-552.
\bibitem{Hartshorne} R. Hartshorne, Algebraic geometry, Springer-Verlag, 1977.
\bibitem{HCS} S. Howard, R. Calderbank, and S. Searle, “A fast reconstruction algo- rithm for deterministic compressive sensing using second order Reed- Muller codes,” Conf. on Info. Sciences and Systems (CISS), Princeton, New Jersey, 2008.
\bibitem{JXC} S. Jafarpour, W. Xu, B. Hassibi and R. Calderbank, Efficient and robust compressed sensing using optimized expander graphs, IEEE Transactions on Information Theory,  Vol. 55 (2009), no.9, 4299-4308.
\bibitem{HK} M.Homma and S.J.Kim, An elementary bound for the number of rational points of a hypersurface over finite fields, Finite Fields and Their Applications, Vol.20 (2014), 76-83.
\bibitem{LG} S. Li, F.Gao, G.Ge and S.Zhang, Deterministic construction of compressed sensing matrices via algebraic curves, IEEE Transactions on Information Theory, Vol.58 (2012), no.8, 5035-5041.
\bibitem{Mohads} M.M. Mohads, A. Mohades and A.Tadaion, A Reed-Solomon code based measurement matrix with small conherence, IEEE Signal Processing Letter, Vol.21 (2014), no.7, 839-843.
\bibitem{NT} J. Nelson and V. N. Temlyakov, On the size of inherence system,  J. Approx. Theory, vol.163 (2011), 1238-1245.
\bibitem{Ellip} J.H.Silverman, The arithmetic of elliptic curves, GTM 106, Springer-Verlag, 1986.
 \bibitem{TC} T. Tao and E. J. Cand\'es, Near-optimal signal recovery from random projections: universal encoding strategies?, IEEE Transactions on Information Theory Vol.52 (2006), 5406–5425.
\bibitem{TV} M. A. Tsfasman and S. G. Vladut, Algebraic-geometric codes, Dordrecht, Kluwer, 1991.
\bibitem{YZ} N.Y.Yu and N.Zhao, Deterministic construction of real-valued ternary sensing matrices using optical orthogonal codes, IEEE Signal Processing Letter, Vol.209 (2013), no.11, 1106-1109.




\end{thebibliography}

\end{document}